\documentclass[useAMS,usenatbib]{mn2e}

\newcommand{\msun}{M$_{\sun}$}

\newcommand{\msuns}{M$_{\sun}~$}

\newcommand{\mnras}{MNRAS}
\newcommand{\apj}{ApJ}
\newcommand{\apjl}{ApJ}

\newcommand{\aap}{A\&A}

\newcommand{\nat}{Nature}

\usepackage{graphicx}
\usepackage{amsmath}

\title[Collisional formation of very massive stars in dense clusters]{Collisional formation of very massive stars in dense clusters}
\author[N. Moeckel and C.J. Clarke]{Nickolas Moeckel$^{1}$\thanks{E-mail:
moeckel@ast.cam.ac.uk} and Cathie J. Clarke$^{1}$\\
$^{1}$Institute of Astronomy, University of Cambridge, Madingley Road, Cambridge, CB3 0HA\\
}
\begin{document}

\date{Accepted XXX. Received YYY; in original form ZZZ}

\pagerange{\pageref{firstpage}--\pageref{lastpage}} \pubyear{2009}

\maketitle

\label{firstpage}

\begin{abstract}
We investigate the contraction of accreting protoclusters using an extension of {\it n}-body techniques that incorporates the accretional growth of stars from the gaseous reservoir in which they are embedded.  Following on from Monte Carlo studies by Davis et al., we target our experiments toward populous clusters likely to experience collisions as a result of accretion-driven contraction.  We verify that in less extreme star forming environments, similar to Orion, the stellar density is low enough that collisions are unimportant, but that conditions suitable for stellar collisions are much more easily satisfied in large-{\it n} clusters, i.e. {\it n} $\sim 30,000$ (we argue, however, that the density of the Arches cluster is insufficient for us to expect stellar collisions to have occurred in the cluster's prior evolution).  We find that the character of the collision process is not such that it is a route toward smoothly filling the top end of the mass spectrum.  Instead, runaway growth of one or two extreme objects can occur within less than 1 Myr after accretion is shut off, resulting in a few objects with masses several times the maximum reached by accretion.  The rapid formation of these objects is due to not just the post-formation dynamical evolution of the clusters, but an interplay of dynamics and the accretional growth of the stars.  We find that accretion-driven cluster shrinkage results in a distribution of gas and stars that offsets the disruptive effect of gas expulsion, and we propose that the process can lead to massive binaries and early mass segregation in star clusters.
\end{abstract}


\section{Introduction}
The formation of massive stars via collisions is an idea that has perhaps always been seen as exotic, compared to the less dramatic accretion model.  In the context of clustered star formation, the idea was introduced by \citet{bonnell98a}, who considered the response of a star cluster as it accretes gas from its birth environment.  The idea was further explored in \citet{bonnell02}.  However, the need for collisions as an alternative to accretion became less pressing as modelers of massive star formation pushed into two and three dimensions \citep[e.g.][]{yorke02,krumholz09}, overcoming the radiation-pressure concerns of one-dimensional treatments \citep{wolfire87}.  The success of accretion models, in addition to the extreme stellar densities required for collisions to become viable, pushed most consideration of collisions to the side.

Early studies of collisional star formation tended to take the ONC as their touchstone to reality, a natural choice as the closest region of massive star formation.  Orion by no means represents an extreme of star formation; star clusters can be orders of magnitude more massive.  More recent work has pointed toward these truly massive clusters as sites where accretionaly induced collisions may play a role in stellar growth.  \citet{clarke08} developed analytical arguments, corroborated by later Monte Carlo simulations \citep{davis10}, suggesting that in populous clusters undergoing vigorous accretion, collisions in the core may be significant.  Essentially, these arguments boil down to he fact that a larger-{\it n} cluster, with a longer two-body relaxation time, can be driven to higher densities by accretion before entering core collapse. 

Runaway collisions can occur in these clusters in the absence of star formation considerations, driven by the gravitational dynamics of a system with a mass spectrum attempting to reach equilibrium \citep[the Spitzer instability,][]{spitzer69}.  Studies of this process include both direct {\it n}-body and Monte Carlo calculations \citep[e.g.][]{portegies-zwart02,gurkan04,freitag06}.  A study tailored to the Arches cluster found that collisional runaway could take place on timescales of a few Myr, starting from a full mass function from 0.1--100 \msun in an initially mass-segregated state \citep{chatterjee09}.  It is very young dense clusters like this that we are most interested in.  In this paper, we work toward bridging the gap between hydrodynamic studies of young gas-dominated clusters and purely dynamical studies of runaway collisions, using an {\it n}-body code extended to approximately treat the simultaneous growth and dynamics of a massive embedded cluster.    The space of cluster parameters and accretion scenarios is vast, and we do not attempt a comprehensive study, which will wait for follow-up papers.  Instead we focus on idealizations of two cluster types, one with modest numbers where collisions are not expected, and one more populous where collisions play a more prominent role.

\section{Numerical method and initial conditions}
The details of a star forming region can only be truly explored with multi-physics, cluster-scale calculations.  Currently, the computational cost of these calculations precludes following the clusters out of the gas dynamical phase and into the purely stellar dynamical regime, and a simulation of an Arches-scale cluster seems well over the horizon.  While {\it n}-body codes are much faster, cluster formation is an interplay of gas- and stellar-dynamical processes, and the true applicability of purely gravitational studies in these matters is not clear.  In this work we take a small step from the {\it n}-body side toward the multi-physics camp, by including a simple accretion effect in the cluster dynamics.

The base code for our work is the GPU-enabled version of {\sc NBODY6} \citep{aarseth00,aarseth03}.  We have modified the code so that stars grow in mass over some timescale; in this initial study at a constant rate proportional to their initial mass, so that the shape of the input mass function is preserved.  In our current accretion prescription the gas that accretes onto a star is assumed to be at rest, so that to conserve momentum the stars slow down as they gain mass, and fall into the potential well of the cluster.  In contrast to the studies that motivated this work \citep{clarke08,davis10} the gas that accretes onto the system does not fall from outside the cluster.  Instead, the initial cluster is dominated by a spatially-static potential identified with the natal gas.  This potential is globally lowered at the same rate the stars grow, and is removed on an effectively instantaneous timescale at some point in the simulation, at which point accretion onto the stars halts and the experiment is a pure {\it n}-body one. 

 At all times stars are assumed to have the main-sequence radius appropriate to their mass, which is likely a conservatively small value \citep[especially after a merger of massive stars;][]{suzuki07}.  When collisions are detected the stars merge with no mass loss.  Winds and stellar evolution are not included.  The actual mass of the collision products are thus a strict upper limit.  The efficiency of a collision is realistically less than 100 per-cent \citep[e.g.][]{lombardi96,freitag05}, and the subsequent stellar evolution of a massive star can reduce the accumulated mass drastically, while increasing the radius \citep[see the thorough discussion in][]{glebbeek09}.  A more realistic treatment of collisions would introduce further randomness to the system, which at this stage of these investigations we prefer to avoid.

We consider two types of clusters, focusing mainly on a populous cluster with $n=32768$ (32k), as well as a lower-mass cluster with $n = 2048$ (2k).  The details of the initial conditions are in table \ref{summarytable}.  The stellar initial mass function is a single power law with a Salpeter slope, $\xi(m) \propto m^{-2.35}$, over the mass range 0.03 -- 3.0 \msun.  Over 1 Myr, the stars gain mass so that they span the range 0.3 -- 30.0 \msun, at which point the gas is expelled by fiat.
  All of the simulations presented here are performed with a global star formation efficiency of 30 percent.  The stars and gas are both initially Plummer potentials with matching scale radii.  Initial experiments with alternative density structures do not qualitatively alter the results.  These choices are arbitrary, although plausible; further papers will explore alternatives and enhancements to this basic setup.

\section{32k results}
\begin{figure}
 \includegraphics[width=85mm]{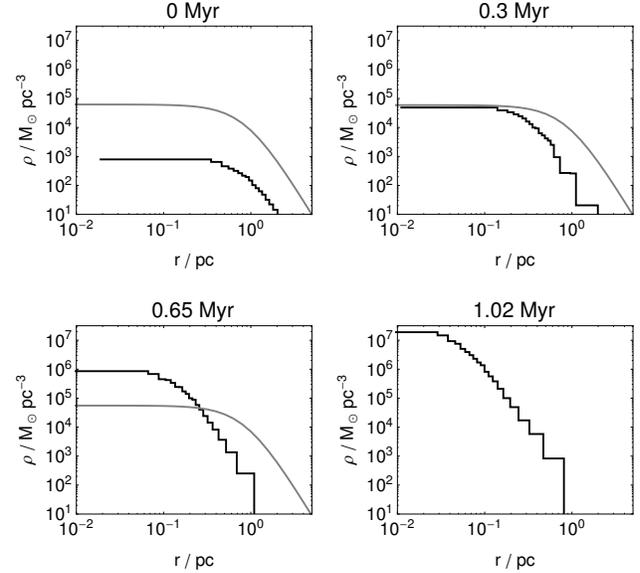}
 \caption{Mass densities for run 32\_2\_a at four times, as labeled.  {\it Black} histograms show the data from the experiment, with stars binned in equal-number bins.  {\it Gray} curves show the value of the gas potential.}
 \label{32_2_a_dens}
\end{figure}

\begin{figure*}
 \includegraphics[width=160mm]{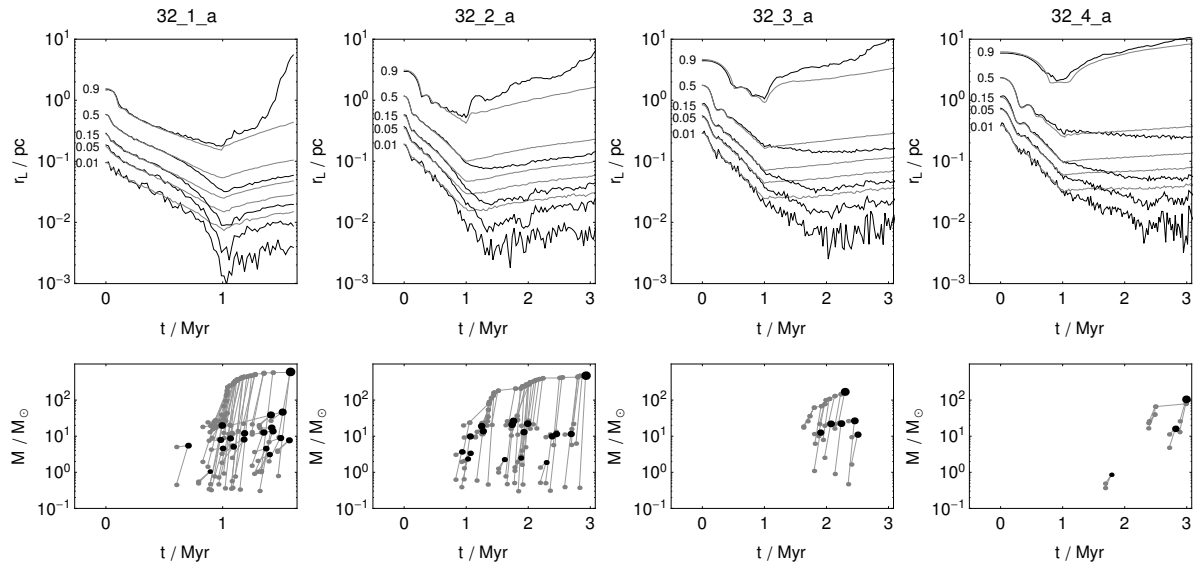}
 \caption{{\it Top row}: Lagrangian radii for clusters with initial virial radii 0.75, 1.5, 2.5, and 3.25 pc,  increasing from the left.  Two stellar populations are shown: in {\it gray}, the lightest third of the total mass; in {\it black}, the most massive third.  The more compact clusters begin to mass segregate during the accretion phase, which ends at 1 Myr.  {\it Bottom row}: the collisional history of the same clusters.  The point at the apex of three points connected by lines is the merger product of of the other two points.  Larger black points are those that exist at the end of the simulation, i.e. they are the end product of a cascade of mergers.}
\label{lagr_rads32}
\end{figure*}

Collisions are expected to occur in the approach to and aftermath of core collapse.  In a single mass system with no gas potential, core collapse occurs via gravothermal contraction.  As energy is transferred from the interior of the cluster outwards via two-body relaxation, the stars in the center slow down, drop further into the potential, and gain energy.  This negative heat capacity leads to a continuous rise in the central density as the cluster's outer radius expands \citep{lynden-bell68}.  As this is a relaxation-driven process, the half-mass relaxation time is the natural timescale to consider \citep{spitzer87}:
\begin{equation}
t_{rh} = 0.138 \frac{n^{1/2}r_h^{3/2}}{(G m)^{1/2}{\rm ln} (\gamma n)},
\end{equation}
where $r_{h}$ is the half-mass radius, $n$ the number of stars, and $m$ the mean stellar mass.  The argument of the Coulomb logarithm depends on the mass spectrum of the stars, and is usually taken in the range $\gamma = $0.01--0.1, with the lower values for a mass-spectrum and the higher value for an equal-mass system \citep{giersz96}.  Core collapse of Plummer sphere with equal masses takes place after about 15 $t_{rh}$.

When a cluster consists of stars with different masses, mass segregation drives the massive stars to the center.  This massive core attempts to achieve energy equipartition with the halo of lower-mass stars.  \citet{spitzer69} showed that for a two-component mass spectrum, the ability of the core of massive stars to achieve energy equipartition with the lower-mass halo depends on the relative masses and total mass of the two components.  For realistic mass functions, the transfer of energy from the massive core to the halo actually drives the system further out of equipartition, and core collapse ensues.  This process, the Spitzer instability, generalizes to a mass spectrum instead of a two-component system \citep[e.g.][]{gurkan04}.  Core collapse with a broad mass spectrum generally occurs at $t\le 0.1 t_{rh}$, much shorter than for the single-mass system.

When the gas potential dominates the stellar potential, the radial scaling of the stellar cluster scales with the accumulated mass as  $r \propto M^{-1}$, and the usual timescales for relaxation effects are complicated by the availability of a potential besides that of the collisional bodies.  If the stellar potential begins to dominate a region, the accretion scenario is more akin to mass flowing onto a core from the outside, and the radius-mass scaling turns over to scale as $r \propto M^{-3}$ \citep[these scalings are discussed in][]{bonnell98a}.  At this point familiar relaxation effects and the Spitzer instability are able to act, and the massive stars may decouple from the low mass stars and begin to go into core collapse.  In figure \ref{32_2_a_dens} we show an example of the density evolution of the stars relative to the gas for one of our runs.  Initially the gas dominates the stellar mass at all radii.  As the stars gain mass and migrate inwards, the stellar density increases toward the center, eventually dominating the central potential.

The behavior of our accreting clusters can be broadly divided into two regimes, depending on the initial conditions.  In the first, the relaxation time of the core becomes short enough that the Spitzer instability can act prior to gas expulsion, so that both accretionaly driven contraction and dynamical effects drive the system rapidly to core collapse.  In the second regime, the high and low mass stars remain coupled throughout accretion, which drives the contraction. When the gas is expelled the central regions are minimally affected, quickly revirialize, and the Spitzer instability alone completes the collapse.  With our fixed mass accretion rate, the initial radius of the cluster determines which path the cluster evolution takes.  

In figure \ref{lagr_rads32} we show the Lagrangian radii (Radii enclosing fixed mass fractions) for each of the four initial cluster radii that we examined.  The two clusters on the left, 32\_1\_a and  32\_2\_a, are in the first regime where mass segregation has begun at the time of gas expulsion at 1 Myr.  In the two larger-radii experiments, the mass segregation leading to core collapse commences only after the gas is expelled.  We performed five realizations each at the cluster radii just on each side of the divide (series 32\_2 and 32\_3), to ensure that the qualitatively different collision behavior in these regimes is a general trend and not a quirk of a single run.  We performed just a single run at each of the two extremes of our initial radius range, as we use them just to illustrate the trends in the bulk cluster properties, which vary little between realizations.  The initial conditions and results are found in table \ref{summarytable}.  All the models were run for 3 Myr, except for run 32\_1\_a.  This run became numerically problematic when the most massive star exceeded 600 \msun.  By this point the simplifying assumptions we make regarding collisions have been stretched to their limits, and we stopped the run at 1.6 Myr.  The results quoted below are at 2 Myr after the simulation begins, to more clearly differentiate the importance of the accretion process from standard dynamical evolution.

\subsection{Collisional behavior}
The bottom row of figure \ref{lagr_rads32} shows the collision trees for the clusters.  In these plots, the product of a stellar merger is joined by lines to the two stars that took part in the collision.  Collision products that are the end of an individual merger cascade, so are still around at the end of the simulation, are shown as larger black dots.  In all but the largest-radius simulation, a clear runaway collision is seen as a a single star is involved in multiple, closely spaced collisions, rapidly gaining mass.  In all cases, the seed for the collisional runaway is one of the most massive stars in the simulation, which is in a dynamically formed binary.  The run 32\_4\_a the massive stars is formed from just 5 massive stars, which sequentially merge to form a 104 \msuns object.  However, this collisional growth only begins after $\sim 2.25$ Myr.

While some collisions may occur prior to the minimum of the inner Lagrangian radii, in general the collisions begin in earnest once the core has collapsed and binaries have formed dynamically during three-body encounters.
In all cases one or two stars rapidly accumulate mass to become the dominant body in the cluster, reaching masses in excess of 100 \msun.    This process is similar to studies of runaway mergers leading to intermediate mass black holes mentioned in the introduction; the difference here is that this runaway may occur as a result of the cluster's initial growth rather than its later, purely dynamical evolution.  The initial conditions of a standard {\it n}-body simulation would begin at our $t=1$ Myr, when the more compact configurations have already begun the collisional runaway process.  

\citet{clarke08} and \citet{davis10} predicted that the number of collisions in an accreting cluster should be independent of the initial cluster radius, which is clearly not the case in this work.  This difference is because those studies assumed that accretion onto the core continues up through the point of core collapse.  Because of our gas expulsion choice, this is only true for our run 32\_1\_a.  An inspection of figure \ref{lagr_rads32} shows that accretion results in approximately an order of magnitude reduction in the cluster radius, and core collapse gives another order of magnitude for the inner Lagrangian radii, independent of the cluster radius at gas expulsion.  The maximum density (and hence collision rate) in our models therefore depends on the initial radius.  If we had allowed accretion to continue longer for the initially larger clusters, the relaxation time would have become short compared to the accretion time at approximately the same density as the more compact clusters, and the peak density and collision rates would be similar.

We re-emphasize here the idealized nature of the collisions in this work.  There are a number of unmodeled processes that can alter the number of collisions and the final mass of the collision product.  Our decision to take the radius of a merged star as its main sequence radius ignores the initially out-of-equilibrium configuration, including a puffed-up envelope, that results from collisions between main-sequence stars\citep{lombardi96} or pre-main-sequence stars \citep{laycock05}.  An enlarged envelope can increase collision rates and the circularisation of hard binaries, and our collision rates are thus conservative.  The most severe simplification we make, however, is not following the stellar evolution of the collision products.  \citet{glebbeek09} and \citet{chatterjee09} showed, subject to uncertain modeling of stellar evolution at very high masses, that massive stellar winds can prevent the buildup of runaway products to more than about 100 \msun.  The number of collisions per Myr in those authors' simulations were several times less than in these simulations, however, so those results can be taken only as a guide.

\subsection{Effect of binaries}
Dynamically hard binaries can serve to offset core collapse by inflating the cluster core in 3-body interactions.  In that sense they work against a collisional scenario.  However, a binary can itself be perturbed sufficiently to lead to a merger of the two components.  A full study of the effect of binaries requires a vastly expanded parameter space, but to illustrate the different behavior we might expect we repeat run 32\_2\_a with the 328 most massive stars (the most massive 1 percent) replaced by equal-mass binaries (run 32\_2\_a\_b).  The mass function of stellar {\em systems} is then the power-law IMF used in the other experiments, while the individual stellar mass function is different at the high end.  This is not a realistic treatment, but serves to explore the general effects.  The binaries are set up initially on circular, wide orbits; because the binary orbits shrink adiabatically just as the cluster does, we set them to be sub-AU scale at the time when core collapse begins, which places them on the hard side of the hard/soft binary divide by a factor of a few.

The central density at 2 Myr in this run is virtually the same as in the equivalent run with no initial binaries.  The `depth' of core collapse, measured by the minimum over all time of the inner Lagrangian radii, is deeper when binaries must be formed dynamically via three-body interactions.  The cluster core inflates slightly in the latter case, while in the run with binaries the minimum is reached with no overshooting.  The collisions in both cases are dominated by a single object, but in the run with primordial binaries more merger products are formed; 38 compared to 17 at 2 Myr.  The binaries thus appear to act as the seeds of collision, but are no more effective at rapidly inflating the core than the dynamical binaries that form naturally.  

At the moment we do not include a treatment of tidal interactions, which can be an important source of binaries, although the efficiency of the tidal capture process is not fully understood.  At  densities of $\sim 10^7$ pc$^{-3}$, however, rate estimates for three-body capture \citep{goodman93} and tidal capture \citep{press77} indicate that the three-body channel will be more important in these cores.  Another binary formation method is capture by a disc of material, which provides a dissipative environment around a star; the disc can result from tidal disruption of a low-mass star in a close encounter with a massive star \citep{davies06}.  This process is more important when a massive star-disc is involved, although the high velocity dispersion in the core will decrease its effectiveness \citep{moeckel07a}.  While these additional contributions to the binary formation rate could conceivably lead to an earlier onset of collisions, the primordial population is likely to be the most important factor.

\subsection{Response to gas expulsion}
An interesting consequence of the shrinking scale radius of the stars relative to the gas is the robustness of the cluster to gas expulsion.  By driving the stars to the center of the gas potential, the effective star formation efficiency is raised to levels such that the core of the cluster is virtually unaffected by the removal of the gas, even as the global efficiency remains low.  A similar effect was seen (in a smaller-scale cluster) in \citet{moeckel10}, where the long-term dynamical evolution based on the hydrodynamic simulation of \citet{bate09} was less affected by gas removal than a simple estimate based on the global star formation efficiency would suggest \citep[e.g.][]{baumgardt07}.  This is an attractive feature of this model, offering a way to begin the purely dynamical evolution in a very compact configuration despite the removal of the cluster's gas.

The accretion scenario modeled here, with the gaseous background at rest relative to the cluster center of mass, yields the most extreme cluster shrinkage possible barring unrealistic possibilities such as a gas cloud with a counter-rotating stellar population.  Examining figure \ref{32_2_a_dens}, we see that at just 0.3 Myr the inner regions of the cluster have reached an effective star formation efficiency of $\sim 50$ percent, and by 0.65 Myr the gas is negligible in the inner 0.25 pc.  If the efficiency of the mass-loading of the stars were reduced, for example by locally-correlated gas and stellar motion, or the accretion phase represented a smaller fraction of the total mass growth, it still seems likely that the local star formation efficiency in the core will be quite high.

The global star formation efficiency will change the details of this effect, as well as the collisional behavior.  However, these changes should be small.  When the gas potential dominates, the radial scaling of the stars is proportional to their mass as $M^{-1}$, and thus the mass density of stars as $M^{4}$.  This is not changed by the amount of gas that makes up the potential.  An inspection of figure \ref{32_2_a_dens} shows that even with 10 times more gas in the initial cluster, by 0.65 Myr the stars would begin to dominate the central potential.  The time of core collapse would be pushed later, and since core collapse would begin from a more compact configuration, the peak density would be greater.  As long as the stars reach the point of dominance, the resistance of the clusters to dissipation from gas expulsion should be largely insensitive to the global efficiency.

\subsection{Density profiles: comparison to the Arches}
\begin{figure*}
 \includegraphics[width=160mm]{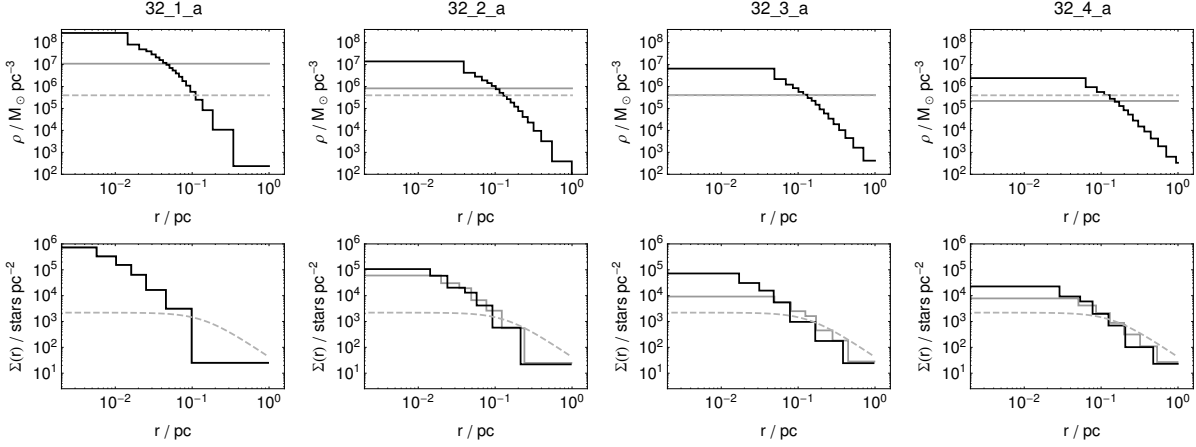}
 \caption{{\it Top row}: volume density profiles at 2 Myr for the same runs shown in figure \ref{lagr_rads32}.  The dashed horizontal line is the cluster density of the Arches found by \citet{figer99}, and the solid horizontal line is the analogous density from our simulations.  {\it Bottom row}: surface density of the 535 most massive stars in the simulation, for comparison to the empirical fit to the Arches from \citet{espinoza09}, shown as a dashed curve.  The {\it gray} histogram is at 1 Myr, the time of gas expulsion.  The {\it black} histogram is at 2 Myr.}
\label{densities32}
\end{figure*}
The present day Arches cluster is notable for its large mass and compact configuration.  As the densest known cluster in the Galaxy, this is the natural place to look for the effects we are considering here.  Mass estimates for the cluster range from $\sim 1.2$--$2\times 10^4$ \msun \citep{figer99,espinoza09}, and the average projected radius of the observed stars is 0.19 pc \citep{serabyn98,figer99b}; this leads to a central density estimate of $\rho_c \geq 3\times10^5$ \msuns pc$^{-3}$.  

In figure \ref{densities32} we plot  density profiles at 2 Myr for runs 32f.  For these plots we consider only stars within 1 pc of the density center, roughly the present-day tidal radius of the Arches.  The peak densities in our models are of the order $10^6$--$10^7$ \msuns pc$^{-3}$, higher than the estimated density of $\geq 3 \times 10^5$ \msuns pc$^{-3}$.  This density estimate is based on the total mass of the cluster as extrapolated from counts of massive stars and a cluster radius given by the mean projected separation of each star from the centroid.  We can form an analogous density estimate for our models, by taking the total mass and the average radius of each star.  This value, as well as the observed estimate, are plotted as solid and dashed gray lines respectively.  The agreement there is much better than comparing our peak density, although the uncertainty in the cluster mass relative to our model ($M_{tot}\sim3\times10^4$\msun) should be kept in mind.  This mean density and radius are given as $\rho_m$ and $r_m$ in table \ref{summarytable}.

While the estimates of the Arches mass density are limited to a mean value, 
\citet{espinoza09} found an empirical surface density profile using stars in the range $10 \le M/$\msun$\le 120$, using a King surface density profile, 
\begin{equation}
\Sigma(r) = \frac{\Sigma_0}{1+(r/r_c)^2}.
\end{equation}
Their best fit for the Arches takes the values $\Sigma_0 = 2.2\times10^3$ stars pc$^{-2}$, and $r_c = 0.14$ pc.  This fit contains $\sim 535$ stars within 1 pc.  While we cannot use the same mass range because of our arbitrary cutoff of 30 \msun, we can construct an analogous surface density profile with the most massive 535 stars in our experiment.  This is also shown in figure \ref{densities32}.  The surface density of the massive stars in our model is clearly more centrally concentrated than the actual cluster.

We are left with the following situation: in order to induce core collapse and collisional runaway before (or very shortly after) the accretion phase of the cluster, for reasonable accretion timescales the density of the stars at the age of the Arches is far too high to match observations.  Cluster setups that more closely match the density profiles never begin mass segregation during the accretion phase; after the expulsion of the gas, the cluster is similar to standard initial conditions for a pure {\it n}-body simulation \citep[e.g.][]{chatterjee09,harfst09}.  Accretion driven collapse as modeled here may have contributed to the compact observed conditions of the Arches, which could lead to subsequent core collapse and collisions as argued by \citet{chatterjee09}.  However, our simulations show that it is unlikely that the Arches was ever dense enough in the {\em past} to have already undergone a collisional runaway.  The run that most closely matches the Arches, 32\_4\_a, still has a surface density several times higher than observed, and it experienced only a single collision in the first 2 Myr.

\section{2k results}
\begin{figure}
 \includegraphics[width=80mm]{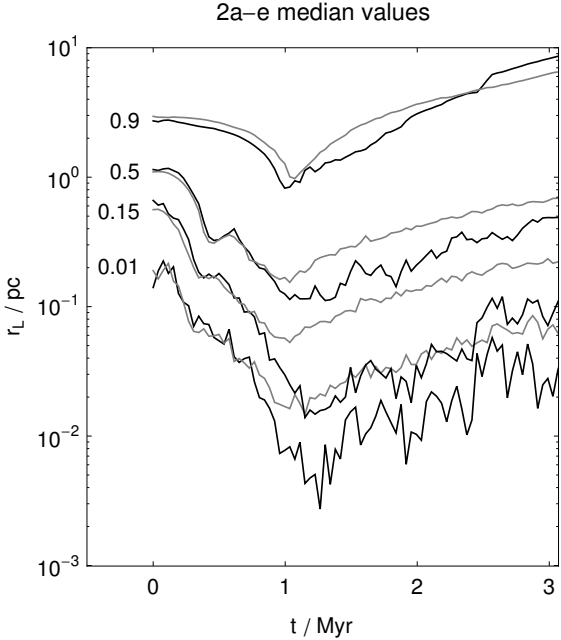}
 \caption{Lagrangian radii for the 2k cluster, shown as the median value of the five runs.  In {\it gray}, the lightest third of the total mass; in {\it black}, the most massive third  }
 \label{onclagrangerads}
\end{figure}
Previous work suggests that collisions should be unimportant at the number of stars and density of Orion \citep{bonnell98a,clarke08,davis10}.  As a check that our scenario does not result in significant collisions in such environments, and is able to yield a cluster with similar bulk properties to the ONC, we performed five realizations of a cluster with 2048 stars tailored to end up as an ONC-like cluster.  The initial conditions and summary of the results for each of these runs are found in table \ref{summarytable}.  The results quoted below and in the table are at 3 Myr, roughly the age of the ONC.

\subsection{Collisional behavior}
In two of the five runs a single collision product is formed.  In 2b, a 14 \msuns and a 6 \msuns star merge, and this product later collides with the most massive star in the cluster.  In 2e, the two most massive stars merge after forming a binary and the product later collides with a 0.3 \msuns star.  All of the collisions result from perturbed, dynamically-formed binaries in the very central regions of the cluster.

\subsection{Density evolution}
In figure \ref{onclagrangerads} we show the Lagrangian radii of the 2k runs.  Because of the noisiness of these data (due to the lower number of stars) we plot the median value of the five runs.
At the time of gas expulsion the mass in the center of the cluster is dominated by the stars, just as for the 32k cases.  The initial radius for these clusters is close to the 32\_2 series, but because of the lower number of stars the relaxation time is a factor of $\sim 2$ smaller, and the runs are slightly further into Spitzer instability-driven core collapse before gas expulsion.  

The 2k clusters expands more after gas expulsion relative to the 32k runs; nearly an order of magnitude expansion of the innermost Lagrangian radii, compared to a factor of $\sim 2$ for the 32\_2 series.  This is not due to the removal of the gas, but rather the increased effect of a few hard binaries in the core when $n$ is small.  This result reinforces the arguments of \citet{clarke08} that stellar dynamical relaxation effects make the production and maintenance of high density cluster cores more difficult in low-{\it n} systems.  At 3 Myr the central densities and half-mass radii of these models compare well to the ONC's values of 2--3$\times 10^4$ \msuns pc$^{-3}$ and $\sim$0.8 pc \citep{hillenbrand98}.  Just as for the 32k simulations the details of the final profile depend on the initial distribution, but this suffices to show that an ONC-like cluster can be created in this model without an alarming number of collisions.  

We also note that the result of the early core collapse in each case is a few binaries containing massive stars, and some degree of mass segregation (visible in the Lagrangian radii).  It is possible that an accretion-driven core collapse process similar to the one modeled here could help resolve the mismatch between the apparent dynamical age of the ONC and its actual age.  An early dense phase in Orion's history has been proposed by \citet{allison09b,allison10} as a way to form the Trapezium.  Those authors invoke cold, fractal, gas-free initial conditions to drive the magnitude of the cluster's potential energy to high values, allowing collapse to a very compact configuration.  The accretion scenario explored here provides an alternative route to a similar result.

\section{Discussion}
The experiments presented here by no means constitute a complete examination of the importance of collisions during the accretion-driven contraction of forming clusters.  The Plummer spheres used for the initial conditions and the highly idealized accretion model  serve instead to illustrate the basic process.  In future papers we will explore the more subtle aspects of this scenario, by adopting more realistic collision, accretion, and stellar evolution prescriptions, varying the density profiles of the stars and the gas potential, and investigating the effect of binaries much more thoroughly. 

We find that the density profiles resulting from any cluster which hosts a significant number of collisions are too centrally concentrated to match known Galactic clusters.
While the single-valued density estimate made in analogy to observations is of the right order of magnitude compared to the Arches, the surface density is too high.   The details of the density profile depend on the initial conditions, and a more sophisticated treatment of several processes including collision behavior, stellar evolution, and the behavior of the gas in the simulations.  Recall that the gas potential is fixed in radius and instantaneously removed.  This choice was made for simplicity, and to avoid the temptation to massage the gas expulsion details to result in a particular density profile.     Recently there has been some interest in fractal or otherwise clumpy initial conditions, particularly in the context of early mass segregation \citep{mcmillan07,moeckel09b,allison09b,allison10}.  The effects seen in those papers may accelerate the concentration of stars in the center of the cluster, potentially impacting the collision processes seen here.  However since none of those works include any gas, which is almost totally dominant gravitationally during the formation stage we are exploring here, this is pure conjecture.  In longer term ongoing simulations, we are examining these factors in much greater detail.  

The basic results of these simulations seem to be robust.  In all large-{\it n} simulations that begin mass segregation prior to the end of accretion, many collisions occur, resulting in a runaway object with a final mass well in excess of the maximum reached through accretion, 30 \msuns in these models.  In cases where mass segregation begins only after gas expulsion, the cluster evolution proceeds to core collapse as studied by previous authors.  In these cases, their initial conditions correspond to the middle of our simulations.  The tendency of one or two objects to dominate the collision process is important.  This route is evidently not a means to fully populate the upper end of the mass function, but rather a means toward generating some of the most extreme objects.  

The similarity to previous work on the runaway growth of IMBH progenitors is clear.  In this case, however, the core collapse that leads to runaway collisions is not necessarily a result of dynamics alone, starting from a cluster with a fully mature IMF.  In our models, core collapse may be driven by the growth of the cluster.  At the point when the natal gas is expelled the clusters are, depending on their initial radius, either already in core collapse or much closer to it than a typical equilibrium model of, for example, the Arches.  As a result, collisions and runway commence on a very short timescale, usually within a few $10^5$ yr.  Importantly, the stars' characteristic radius shrinks relative to the gas potential, increasing the effective star formation efficiency and leaving the cluster resilient to disruption from gas dispersal.  Collisional matters aside, this presents a mechanism to create and maintain the dense initial conditions that a standard {\it n}-body simulation of a dense young cluster might start from.

Finally, we note that it is reassuring that this process, when tailored to reproduce a cluster with similar bulk properties to the ONC, does not result in a significant number of collisions, despite going into core collapse prior to the expulsion of gas.  This result is due to the different behavior of low- and high-$n$ systems when a few hard binaries form at core collapse.  The fact that collisions are expected to be unimportant in Orion, the nearest and best studied site of massive star formation, is evidence against the need for collisions in order to create a typical O or B star.  However Orion, much like current high-mass star formation simulations, has no single system in excess of $\sim 50$ \msun.  It is the Universe's more extreme objects that may require more dramatic formation mechanisms.

\section*{Acknowledgments}
Our thanks to Sverre Aarseth for help debugging our modifications to his code and for letting us use his GPUs, and to the referee for a careful reading and constructive suggestions.

\begin{table*}
 \centering
 \begin{minipage}{150mm}
  \caption{Parameters and results of the experiments.  Results are at 2 Myr for the 32k runs, and 3 Myr for the 2k runs unless noted otherwise.
  }
  \label{summarytable}
  \begin{tabular}{@{}lcccccccc@{}}
  \hline
Run name & $n$ & $r_{V,0}$/pc & $\rho_{c}$/\msuns pc$^{-3}$ & $r_{hm}$/pc & $\rho_{m}$/\msuns pc$^{-3}$ & $r_{m}$/pc & $N_{col}$ & $M > 30$ \msun\\  
    \hline 
 32\_1\_a (at 1.6 Myr) & 32768 & 0.75 & $5.1\times10^7$ & 0.08 & $2.5\times10^6$ & 0.14 & 57 & 602, 46, 39  \\
    \hline    
 32\_2\_a & 32768 & 1.5 & $1.2\times10^7$ & 0.14 & $8.0\times10^5$ & 0.19 & 33 & 276, 37  \\
 32\_2\_b & 32768 & 1.5 & $1.6\times10^7$ & 0.13 & $8.1\times10^5$ & 0.20 & 29 & 344   \\
32\_2\_c & 32768 & 1.5 & $1.3\times10^7$ & 0.14 & $8.1\times10^5$ & 0.20 & 41 & 356, 31  \\
32\_2\_d & 32768 & 1.5 & $1.4\times10^7$ & 0.13 & $8.0\times10^5$ & 0.20 & 26 & 368  \\
32\_2\_e & 32768 & 1.5 & $1.3\times10^7$ & 0.14 & $7.9\times10^5$ & 0.20 & 26 & 321  \\
32\_2\_a\_b & 33096 & 1.5 & $1.3\times10^7$ & 0.14 & $8.3\times10^5$ & 0.20 & 80 & 283, 46, 35  \\         
    \hline 
     32\_3\_a & 32768 & 2.25 & $6.5\times10^6$ & 0.19 & $4.2\times10^5$ & 0.24 & 7 & 100  \\
 32\_3\_b & 32768 & 2.25 & $5.9\times10^6$ & 0.19 & $4.0\times10^5$ & 0.25 & 10 & 143   \\
 32\_3\_c & 32768 & 2.25 & $6.0\times10^6$ & 0.19 & $4.2\times10^5$ & 0.24 & 11 & 145  \\
 32\_3\_d & 32768 & 2.25 & $6.6\times10^6$ & 0.19 & $4.3\times10^5$ & 0.24 & 6 & 65, 45  \\
 32\_3\_e & 32768 & 2.25 & $7.1\times10^6$ & 0.19 & $4.1\times10^5$ & 0.24 & 4 & 57, 39  \\
        \hline   
 32\_4\_a & 32768 & 3.0 & $2.5\times10^6$ & 0.29 & $2.1\times10^5$ & 0.29 & 1 & -  \\
          
        \hline      
2a & 2048 & 1.46 & $4.5\times10^3$ & 0.68 & & & 0 & - \\
2b & 2048 & 1.46 &  $2.5\times10^4$ & 0.65 & & & 2 & 50 \\
2c & 2048 & 1.46 &  $1.7\times10^4$ & 0.83 & & & 0 & - \\
2d & 2048 & 1.46 &  $2.9\times10^4$ & 0.58 & & & 0 & - \\
2e & 2048 & 1.46 &  $2.4\times10^4$ & 0.58 & & & 2 &  55 \\
        \hline     

\end{tabular}
$N$: number of stars in the simulation.\\
$r_{V,0}$: initial virial radius.\\
$\rho_{c}$: central density.\\
$r_{hm}$: half-mass radius.\\
$\rho_{m}$: for the 32k clusters, average density determined by the total mass and mean radius of stars within 1 pc.\\
$r_{m}$: for the 32k clusters, mean radius of stars within 1 pc.\\
$N_{col}$: number of stellar collisions.\\
$M > 30$ \msun: the masses of all stars above 30 \msun.
\end{minipage}
\end{table*}

\bsp

\label{lastpage}

\end{document}